\documentclass[epj]{webofc}
\usepackage[varg]{txfonts}   

\newcommand{\be}{\begin{eqnarray}}
\newcommand{\ee}{\end{eqnarray}}
\newcommand{\ba}{\begin{array}}
\newcommand{\ea}{\end{array}}

\newcommand{\bi}{\begin{itemize}}
\newcommand{\ei}{\end{itemize}}
\usepackage[hypertex,colorlinks=true,linkcolor=red,citecolor=blue]{hyperref}

\woctitle{QUARKS-2016 19th International Seminar on High Energy Physics}
\begin{document}
\title{On the equivalence of GPD representations}
%
%

\author{\firstname{Dieter} \lastname{M\"{u}ller}\inst{1}\fnsep\thanks{\email{dieter.muller@irb.hr}} \and
        \firstname{Kirill} \lastname{Semenov-Tian-Shansky}\inst{2}\fnsep\thanks{\email{cyrstsh@gmail.com}} 
}

\institute{Theoretical Physics Division, Rudjer Bo\v{s}kovi\'{c} Institute, HR-10002 Zagreb, Croatia
\and
Petersburg Nuclear Physics Institute,
Gatchina, 188300, St.Petersburg, Russia          }

\abstract{%
Phenomenological  representations of generalized parton distributions
(GPDs) implementing the non-trivial field theoretical requirements
are employed in the present day strategies for extracting
of hadron structure information encoded in GPDs from the observables of hard exclusive reactions. Showing out the equivalence of various GPD representations can help
to get more insight into GPD properties and allow to build up flexible GPD models
capable of satisfactory description of the whole set of available experimental data.
Below we review the mathematical aspects of establishing equivalence between the
the double partial wave expansion of GPDs in
the conformal partial waves and in the $t$-channel
${\rm SO}(3)$
partial waves and the double distribution representation of GPDs.
}
\maketitle
\section{Introduction}
\label{intro}

One of the the prominent goals of the present day hadron physics is the
extraction of generalized  parton  distributions
(GPDs) (see Refs.~\cite{Diehl:2003ny_semenov,Belitsky:2005qn_semenov,Mueller:2014hsa_semenov} for reviews). These non-perturbative quantities  provide a detailed description of hadronic structure in terms of QCD quark and gluon degrees of freedom.
GPDs can be accessed in hard exclusive reactions
such as the deeply virtual Compton scattering (DVCS) 
(see Ref.~\cite{d'Hose:2016prj_semenov} for a recent review) and hard meson electroproduction off hadrons (HMP) (see Ref.~\cite{Favart:2015umi_semenov}).

The realistic strategy for extracting GPDs from the data relies on employing of phenomenologically motivated GPD representations and fitting
procedures for the whole set of observable quantities (see {\it e.g.}
Ref.~\cite{Kumericki:2016ehc_semenov} and references therein).
The non-trivial requirements (forward limit, polynomiality and support properties, hermiticity, positivity
constraints, evolution properties, {\it etc.}) following from the fundamental properties of the underlying quantum field theory, provide a clue to build up a consistent phenomenological representation for GPDs.

All GPD representations
present the same field theoretical object, and therefore, as the basic theoretical requirements are satisfied, should be equivalent.
Moreover, comparing the manifestation of GPD properties within different representations
could provide additional insight of GPDs and their physical
interpretation. However, sometimes explicitly working out the mapping of a GPD within one
representation to that in a different representation could be
mathematically cumbersome.

Below we present an overview of the current status of equivalence studies between the
double distribution representation of GPDs and the representations based on the expansion of GPDs in conformal partial waves.

\section{Double distribution representation of GPDs and
the conformal partial wave expansion}
\label{Sec_ConfPWvsDD}

\subsection{Double distribution representation}
The double distribution (DD) representation of GPDs introduced in \cite{Mueller:1998fv_semenov,Radyushkin:1997ki_semenov}
was historically one of the first parametrizations for GPDs suitable for phenomenological applications. It arises directly from  the diagrammatical considerations and inherits most of the basic field theoretic requirements by presenting a GPD as a one-dimensional section
of a two-varibale double distribution. It provides an elegant way to implement both the polynomiality, the support properties and the forward limit constraint.
The DD representation is known to be not uniquely defined. However, by means of a `gauge' transformation
\cite{Teryaev:2001qm_semenov}
it can be put into the conventional  form known as the ``DD$+D$'' `gauge' with the
double distribution part supplemented by the so-called $D$-term addendum
\cite{Polyakov:1999gs_semenov}.
For the case of a quark GPD $H$ with the support
$x \in [-\eta;\,1]$
it reads
\be
H^q(x\ge -\eta,\eta,t)=\int_{0}^1 dy \int_{-1+y}^{1-y} dz \,
\delta(x-y-\eta z) f(y,z,t)+ D(x/|\eta|,t),
\label{DD+Dterm}
\ee
where
$x$,
the skewness parameter
$\eta$
and the momentum transfer squared
$t$
stand for the usual GPD variables.

As first pointed out in
\cite{Teryaev:2001qm_semenov},
the relation between GPD and the corresponding double distribution
is a particular case of the Radon transform.
This transformation can be inverted
in order to recover DDs from GPDs by the standard means of the
Radon tomography \cite{Teryaev:2001qm_semenov,Radyushkin:1998bz_semenov}:
\be
f(y,z)=\frac{-1}{2 \pi^2} \int_{-\infty}^\infty d  \eta \, {\rm PV} \!
\int_{-\infty}^\infty \frac{dx}{x} H_{\rm DD}(x+y+z \eta, \eta),
\label{InvRadon}
\ee
where
$H_{\rm DD}=H-D$  stands for the DD part of (\ref{DD+Dterm}) and
`PV' denotes the principal value regularization prescription.
Note that the inversion
(\ref{InvRadon})
requires the knowledge of GPD for
$|\eta|>1$
which is given by the cross channel counterpart of GPD, the so-called
generalized distribution amplitude (GDA):
\be
&& H(x,\eta,t)
\nonumber \\ 
&& =  \theta(1-|\eta|)H_{\rm GPD}(x,\eta,t)+
\theta(\eta-1) \frac{1}{\eta} H_{\rm GDA}\left( \frac{x}{\eta} ,\frac{1}{\eta},t \right)
 +
\theta(-\eta-1) \frac{1}{\eta} H_{\rm GDA}\left( \frac{x}{\eta} ,\frac{-1}{\eta},t \right).
\ee
The example of the application of the inverse Radon transform 
to recover the double distributions  from GPDs  for the toy model 
case of photon GPDs 
\cite{Friot:2006mm_semenov}
is presented in 
Ref.~\cite{Gabdrakhmanov:2012aa_semenov}. 
However, the mathematical subtleties encountered 
while applying the inversion 
(\ref{InvRadon}) 
for the realistic GPD models deserve further investigation.

\subsection{Conformal partial wave expansion of GPDs and the dual parametrization}
\label{SSec_ConfPW}
The alternative strategy for building up a GPD representation resides on the expansion of GPDs over the conformal partial wave (PW) basis, which ensures the
diagonalization of the leading order (LO) evolution operator.
The conformal PW expansion of a quark GPD $H^q$ is formally given by
\begin{eqnarray}
\label{H(x,eta,t)-CPWE}
H^q(x\ge -\eta,\eta,t) = \sum_{n=0}^\infty (-1)^n p_n(x,\eta) H_n(\eta,t); \ \ \
H_n(\eta,t)=\int_{-1}^1 dx \eta^n c_n(x/\eta) H (x,\eta,t),
\end{eqnarray}
where we adopt the notations of Ref.~\cite{Mueller:2005ed_semenov}. The conformal moments
$H_n(\eta,t)$
are formed with respect to the Gegenbauer polynomials
$c_n(x)=N_n C_n^{\frac{3}{2}}(x) \equiv \frac{\Gamma(3/2) \Gamma(n+1)}{ 2^n \Gamma(n+\frac{3}{2})} C_n^{\frac{3}{2}}(x)$ and
the conformal PWs
$p_n$
also are the suitably normalized Gegenbauer polynomials dotted by the weight
$(1-x^2/\eta^2)$
and the support
$\theta$-function.
The expansion
(\ref{H(x,eta,t)-CPWE})
is to be understood as an ill-defined sum of generalized functions and
require proper resummation in order to be defined rigourously.

Two different receipts were proposed in the literature to handle the conformal
PW expansion
(\ref{H(x,eta,t)-CPWE}).
The first one
\cite{Mueller:2005ed_semenov,Kirch:2005tt_semenov}
is based on the techniques
of the Mellin-Barnes integral. The resummation of the series is performed
by trading the sum 
in (\ref{H(x,eta,t)-CPWE})
for the Mellin-Barnes integral and the analytic continuation
of PWs and conformal moments to the complex values of the conformal spin.
The second way
consists in the use of the so-called Shuvaev-Noritzsch transform \cite{Shuvaev:1999fm_semenov,Noritzsch:2000pr_semenov}.
In this case one introduces the forward-like functions of an auxiliary momentum-fraction-type variable whose Mellin moments generate the conformal
moments of GPDs. GPD can then be expressed as convolution of the forward-like
functions with certain integral kernels.

In order to achieve further factorization of the GPD variable dependencies
it turns out extremely instructive to further expand the conformal moments over the basis of
the $t$-channel
${\rm SO}(3)$
rotation group partial waves.

In the context of the Shuvaev transform techniques the resulting GPD representation
is known as the dual parametrization of GPDs
\cite{Polyakov:2002wz_semenov,Polyakov:2008aa_semenov,SemenovTianShansky:2010zv_semenov}.
For the case of a spinless target hadron the cross channel
${\rm SO}(3)$
rotation group expansion goes over the usual Legendre polynomials
$P_{l}(\cos \theta_t) \simeq P_l \left( \frac{1}{\eta} \right)$ 
(up to the power suppressed and small target mass corrections):
\be
 N_n^{-1} \frac{(n+1)(n+2)}{2n+3} H_n(\eta,t)=\eta^{n+1} \sum_{l=0}^{n+1}B_{nl}(t) P_l \left( \frac{1}{\eta} \right).
\ee
For the charge even combination of GPDs
$H^{(+)}(x,\eta,t) = {\rm sign}(x)[H^{q}(|x|,\eta,t)+ H^{\bar{q}}(|x|,\eta,t)]$
we then come to the following double PW expansion
\be
\label{H^{(+)}(x,eta,t)-series}
H^{(+)}(x,\eta,t) = 2\sum_{n=1 \atop {\rm odd}}^\infty \sum_{l=0 \atop {\rm even}}^{n+1}    B_{nl}(t)\, \theta(|\eta|-|x|) \left(\! 1-\frac{x^2}{\eta^2}\right) C_n^{3/2}\left(\frac{x}{\eta}\right) P_l\left(\frac{1}{\eta}\right).
\ee
The Mellin transform of the set of the forward-like
functions
$Q_{2\nu}(y,t)$
of an auxiliary variable
$y$
generates the generalized form factors $B_{nl}$:
$
B_{nl}(t)= \int_0^1\! dy\, y^{n} Q_{n+1-l}(y,t).
$
A  GPD can then be represented as a series of integral transformations
of the forward-like functions.
The charge even combination
(\ref{H^{(+)}(x,eta,t)-series})
reads:
\begin{eqnarray}
\label{Q22H}
H^{(+)}(x,\eta,t)&\!\!\! = \!\!\!& \sum_{\nu=0}^\infty \int_0^1\!dy
\left[ K_{2\nu}(x,\eta|y) - K_{2\nu}(-x,\eta|y) \right] y^{2\nu} Q_{2\nu}(y,t)  \,.
\end{eqnarray}
The integral kernels
$K_{2\nu}(x,\eta|y)$
and
$K_{2\nu}(-x,\eta|y)$
are defined non-vanishing for
$-\eta \le x \le 1$
and
$-1 \le x \le \eta$
respectively.
The closed analytic expressions for these kernels were originally calculated by means of Shuvaev's dispersion technique
\cite{Polyakov:2002wz_semenov}
and can be expressed in terms of the elliptic integrals.

The double partial wave expansion
was also implemented within the Mellin-Barnes integral approach.
 In the literature it is usually referred as the
${\rm SO}(3)$
partial wave expansion
\cite{Kumericki:2006xx_semenov}.

The cross-channel
${\rm SO}(3)$-PW
expansion  of the corresponding conformal moments (\ref{H(x,eta,t)-CPWE}) reads
\begin{eqnarray}
\label{H_n(eta,t)-SO(3)}
&&
H_n(\eta,t)= \sum_{\nu=0}^{\frac{n+1}{2},  \; (n/2)} \eta^{2\nu} H_{n,n+1-2\nu}(t) \hat{d}^{n+1-2\nu}(\eta), \ \ \ {\rm for \ \ odd \ \ (even)} \ \ n,
\end{eqnarray}
where the ${\rm SO}(3)$-PWs
in the spinless target hadron case
are expressed by the reduced Wigner
$d^l_{00}$-functions
$
\hat{d}^l_{00}(\eta) =
\frac{\Gamma\!\left(\frac{1}{2}\right) \Gamma(1+J)}{2^J \Gamma\!\left(\frac{1}{2}+J\right)} \eta^l  P_l\!\left(\!\frac{1}{\eta}\!\right)\,.
$

The final expression for the conformal and cross-channel
${\rm SO}(3)$
partial wave expansion of the
quark part of the GPD $H$  within the Mellin-Barnes integral approach reads:
\begin{eqnarray}
\label{nu-expansion}
H(x\ge -\eta,\eta,t) &\!\!\!=\!\!\!& \sum_{\nu=0}^\infty \frac{1}{2i}\int^{c+2\nu+i \infty}_{c+2\nu-i \infty}\!dj\,
\frac{ p_{j}(x,\eta)}{\sin(\pi[j+1])}\, H_{j,j+1-2\nu}(t)\, \eta^{2\nu} \hat{d}^{j+1-2\nu}_{00}(\eta)
\\
&&\!\!\!\!\!
-
\sum_{\nu=1}^\infty
\eta^{2\nu}\,  p_{2\nu-1}(x,\eta) H_{2\nu-1,0}(t)\,.
\nonumber
\end{eqnarray}
The detailed discussion on the analytic continuation of
both the conformal PWs $p_{n}$ and the double partial waves (dPWs)
$H_{n,n+1-2\nu}$
to the complex values
$n=j$
of the conformal spin can be found in Ref.~\cite{Mueller:2005ed_semenov}.
The value of the intercept $c$ of the Mellin Barnes integral contour is
specified by the relative position
of the rightmost pole of the conformal partial wave and that of the
corresponding dPW.

\section{Establishing equivalence of the GPD representations}
\label{Sec_Equiv}

The complete equivalence of the existing GPD representations was
understood already long ago. However, this still has not been demonstrated
in the strict mathematical sense for all variety of GPD representations.
Below we would like to summarize the recent progress in this field.

\begin{figure}[h]
\centering
\includegraphics[width=13cm,clip]{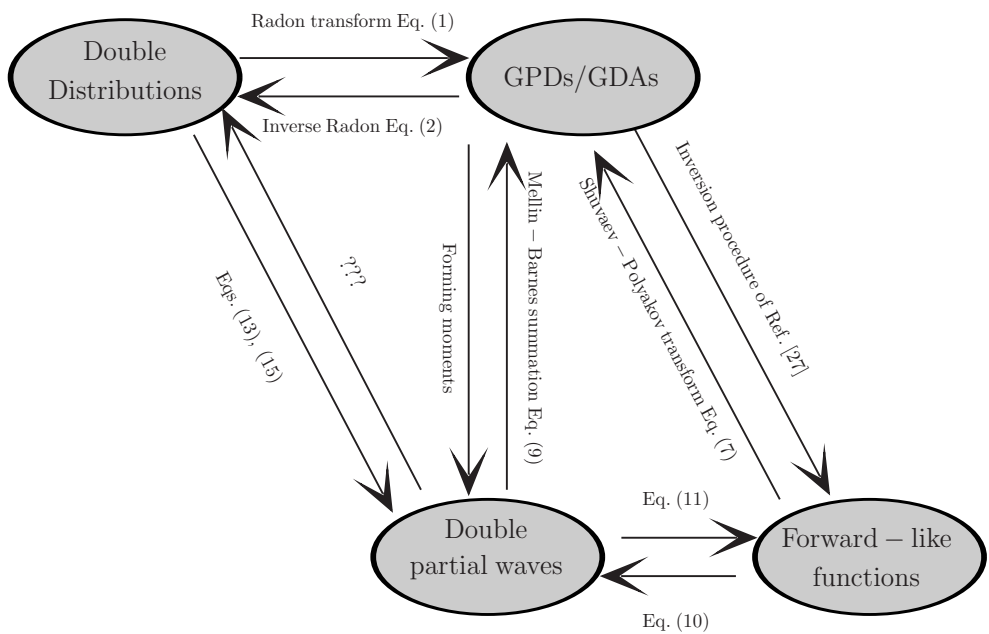}
\caption{Relations between the double distribution parametrization of
GPDs and the two equivalent forms of the double partial wave expansion of GPDs:
${\rm SO}(3)$
partial wave expansion within the
Mellin-Barnes integral approach and the dual parametrization of GPDs. See comments in the text.}
\label{fig-1_semenov}       
\end{figure}

For clearness we show the relation between the DD representation of GPDs and the GPD representations based on the conformal PW expansion in the form of  the ``commutative diagram'' depicted on Fig.~\ref{fig-1_semenov}. We start with the consideration
of the two ``dialects'' of the conformal PW and  the cross channel ${\rm SO}(3)$ PW expansion:
the dual parametrization and the Mellin-Barnes integral technique, specify the dictionary and present the transition rules. Then we comment on the established relation with the DD representation of GPDs.

\subsection{Dual parametrization in two equivalent forms}

The complete equivalence of the dual parametrization of GPDs
and the Mellin-Barnes techniques for the SO$(3)$ conformal partial wave expansion of GPDs
was explicitly established in Ref.~\cite{Muller:2014wxa_semenov}.
The dPWAs
$H_{nl}(t)$
(\ref{H_n(eta,t)-SO(3)}) of the Mellin-Barnes transform approach
can be put in correspondence with the generalized form factors
$B_{nl}(t)$
given by the Mellin moments of the forward-like functions
introduced in the context of the dual parametrization
(\ref{H^{(+)}(x,eta,t)-series}):
\begin{eqnarray}
\label{HjJ2B}
 H^{}_{n,n+1-2\nu}(t) = \frac{\Gamma(3+n)\Gamma\!\left(\frac{3}{2}+n-2\nu\right)}{2^{2\nu} \Gamma\!\left(\frac{5}{2}+n\right)\Gamma(2+n-2\nu)} B_{n,n+1-2\nu}(t).
\end{eqnarray}
The inversion of the corresponding Mellin transform allows to reconstruct the forward-like functions from the dPWAs,
\begin{eqnarray}
\label{HjJ2Q}
y^{2\nu} Q_{2\nu}(y,t) = \frac{1}{2\pi i} \int_{c-i \infty}^{c+i \infty}\!dj\, y^{-j-1}\,\frac{2^{2\nu} \Gamma(5/2+j+2\nu)\Gamma(2+j)} {\Gamma(3+j+2\nu)\Gamma(3/2+j)} H_{j+2\nu,j+1}(t)\,.
\end{eqnarray}

One also can establish the following Mellin-Barnes representation for
the dual parametrization convolution kernels occurring in (\ref{Q22H})
$K_{2\nu}(x,\eta|y)$:
\begin{subequations}
\label{K_{2nu}(x,eta|y)-MB}
\begin{eqnarray}
\label{K_{2nu}(x,eta|y)-MBa}
K_{2\nu}(x,\eta|y) &\!\!\!=\!\!\!& K^{J\neq0}_{2\nu}(x,\eta|y) - \eta^{2\nu}\,  p_{2\nu-1}(x,\eta)
\frac{\Gamma\!\left(\frac{1}{2}\right)\Gamma(2+2\nu)}{2^{2\nu} \Gamma\!\left(\frac{3}{2}+2\nu\right)}\, \frac{1}{y}
\phantom{\Bigg]};
\\
\label{K_{2nu}(x,eta|y)-MBb}
K^{J\neq0}_{2\nu}(x,\eta|y) &\!\!\!=\!\!\!&  \frac{1}{2i}\int^{c+i \infty}_{c-i \infty}\!dj\,  \eta^{2\nu}\,
\frac{p_{j+2\nu}(x,\eta)}{\sin(\pi[1+j])}\, \frac{\Gamma(3+j+2\nu)\Gamma\!\left(\frac{3}{2}+j\right)}{2^{2\nu} \Gamma\!\left(\frac{5}{2}+j+2\nu\right)\Gamma(2+j)}\; y^{j}\,  \hat{d}^{j+1}_{00}(\eta)\,.
\end{eqnarray}
\end{subequations}
Both in 
(\ref{HjJ2Q}) 
and in 
(\ref{K_{2nu}(x,eta|y)-MBb}) 
the contour intercept 
$c$ is chosen as explained in Sec.~\ref{SSec_ConfPW}.
It is straightforward to check that the Mellin-Barnes representation for
kernels
(\ref{K_{2nu}(x,eta|y)-MBb})
leads to the familiar expression for the dual parametrization convolution kernels
\cite{Polyakov:2002wz_semenov,Polyakov:2008aa_semenov}
in terms of elliptic integrals.

As the example of application of the reparametrization procedure
(\ref{HjJ2Q})
the successful
Kumeri{\v c}ki-M{\"u}ller (KM)
model
\cite{Kumericki:2009uq_semenov},
providing global fit to the DVCS world data set for
unpolarized proton target
\cite{Aschenauer:2013qpa_semenov},
was rewritten in terms of the dual parametrization technique.

The established equivalence also contributed much into the clarification
of the $J=0$ fixed pole manifestation issue in DVCS
(see discussion in \cite{Muller:2015vha_semenov}).
It was shown that the
$J=0$
fixed pole universality hypothesis of
Ref.~\cite{Brodsky:2008qu_semenov}
is equivalent to the conjecture that the
$D$-term form factor is given by the inverse moment sum rule for the Compton form factor.
Unfortunately, there exists no theoretical proof for the $J=0$ fixed pole universality conjecture. Therefore, any supplementary $D$-term added to a GPD results in an additional
$J=0$ fixed pole contribution and implies the violation of the 
$J=0$
fixed pole 
universality hypothesis.

\subsection{Mapping DD representation to the conformal PW expansion}

Now we turn to the link between the conformal PW expansion of GPDs and the
the DD representation.
The first example of the reparametrization procedure for this case
was proposed in Ref.~\cite{SemenovTianShansky:2008mp_semenov}.
This procedure is based on the Taylor expansion of GPDs in the vicinity of
$\eta=0$, allowing to map any particular GPD to the forward-like function as it appears in the dual parametrization.
An example of application  was given in
Ref.~\cite{Polyakov:2008aa_semenov}, where
several first forward-like functions reexpressing the Radyshkin double distribution Ansatz   within the dual parametrization approach were computed.

A much more general form of reparametrization procedure linking DD representation
to the double partial wave expansion was constructed in Ref.~\cite{Muller:2014wxa_semenov}.
It consists in the explicit analytic continuation of the corresponding  double partial wave amplitudes to the complex values of conformal spin.

Below for simplicity we quote the result for the GPD within the DD representation
in the DD$+D$ `gauge'
(\ref{DD+Dterm}). The set of dPWAs corresponding to a DD representation with 
a given DD  $f(y,z)$ can be presented in a form 
\begin{eqnarray}
\label{DD2H^{(+)}_{j+2nu,j+1}}
H^{(+)}_{j+2\nu,j+1}
\!\!\!&\!\!\!=\!\!\!&\!\!\! \int_0^1\!dy\!\!\int_{-1+y}^{1-y}\!dz\,
y^{j} f(y,z
)\!\left[
 1+j+  \frac{ y\vec{\partial}}{\partial y}
\right]\!
 {\cal K}_{j+2\nu,j+1}^{-1}(y,z)
+ \delta_{j+1,0}\, D_{2\nu-1,0}(t)\,,
\end{eqnarray}
where the integral kernel
turns to be a polynomial in
$y$
and
$z$.
It can be expressed  through  action of the appropriate differential operator on
the Gegenbauer polynomials.
\begin{eqnarray}
\label{ K_{j+2nu,j}^{-1}(y,z)-1}
{\cal K}_{j+2\nu,j+1}^{-1}(y,z) &\!\!\!=\!\!\!&
\frac{(1+2 \nu)_j (2+j)_\nu}{\Gamma(2+j)\left(\frac{3}{2}+j+\nu\right)_\nu}
\nonumber\\
&&\times\int_{-1}^{1}\!d\omega\,
\frac{\left(2+j\right)_{2+j} \left(1-\omega^2\right)^{1+j}}{2^{3+2 j}\,\Gamma(2+j)}\,
{_2F_1}\!\!\left({-2\nu,3+2j+2\nu\atop 2+j }\bigg|\frac{1-\omega y-z}{2}\!\right).
\nonumber\\
\end{eqnarray}
The closed form of the kernel
${\cal K}_{j+2\nu,j+1}^{-1}(y,z)$
can be worked out in terms of the rather cumbersome Appell hypergeometric function ${F_4}$,
which finally reduces to the finite double sum
\be
{\cal K}_{j+2\nu,j+1}^{-1}(y,z) &\!\!\!=\!\!\!&
\frac{(-1)^\nu 2^{-2 \nu} (2+j)_{2 \nu -1}}{\Gamma(\nu+1) \left(\frac{3}{2}+j+\nu\right)_\nu}
{F_4}(-\nu,\frac{3}{2}+j+\nu,\frac{5}{2}+j,\frac{1}{2},y^2,z^2)
\nonumber \\
&\!\!\!=\!\!\!&
\frac{ (2+j)_{2 \nu -1}}{\left(\frac{3}{2}+j+\nu\right)_\nu}
\sum_{m=0}^\nu \sum_{n=0}^{\nu-m} \frac{(-1)^{\nu-m-n}\,  y^{2m} z^{2n}}{m! n! (\nu -m-n)!\,  2^{2\nu}}\,
\frac{ \left(\frac{3}{2} + j+\nu\right)_{m+n}
}{
\left(\frac{5}{2} + j\right)_{m}\left(\frac{1}{2}\right)_{n}},
\label{K_{j+2nu,j}^{-1}(y,z)-sum}
\ee
where
$(k)_l \equiv \frac{\Gamma(k+l)}{\Gamma(l)}$
stand for the Pochhammer symbol.

The transformation
(\ref{DD2H^{(+)}_{j+2nu,j+1}}), (\ref{ K_{j+2nu,j}^{-1}(y,z)-1})
provides the analytic continuation of the GPD moments in
$j$ for the regular part of the GPD.
The
$j=-1$
Kronecker delta contribution induced by the $D$-term part
can be calculated from a
$\delta(y)$-proportional addenda
to DD
by the convolution with the kernel
$z \frac{\partial}{\partial z} K^{-1}_{2\nu-1,0}(y=0,z)$,
which  explicitly  reads as
\begin{eqnarray}
D_{2\nu-1,0}(t) &\!\!\! =\!\!\!&
\frac{\Gamma\!\left(\frac{1}{2}\right)\Gamma(2+2\nu)}{2^{2\nu} \Gamma\!\left(\frac{1}{2}+2\nu\right)}
\int_{-1}^1\! dz\,  D(z,t)\,\, {_2F_1} \!\left({1-2\nu,2+2\nu\atop 2 }\bigg|\frac{1-z}{2}\!\right).
\end{eqnarray}

\section{Conclusions and outlook}
\label{Sec_Conc_semenov}
Thus all but one equivalence relations depicted on the
``commutative diagram'' Fig.~\ref{fig-1_semenov}
have been constructed explicitly.
The dual parametrization approach is found to be completely
equivalent to the Mellin-Barnes type
integral based techniques for GPDs.
Also the general transformation allowing to recast the DD representation
in the form of double partial wave expansion is proposed.

The only so far missing equivalence relation of Fig.~\ref{fig-1_semenov} is the inverse
relation connecting DPWAs of the double partial wave expansion
(\ref{H_n(eta,t)-SO(3)})
and the double distributions. It can be designed as the proper
combination of the Mellin-Barnes summation procedure and the inverse Radon
transform (\ref{InvRadon}).
The corresponding studies are underway and the result
will be published elsewhere
\cite{SemenovTianShanskyXXX_semenov}.

In particular, it would be extremely instructive to
map the successful KM GPD model into the double distribution representation.
This can lead to lifting off the unjustified ``ad hoc'' assumptions in DD modeling
and hence to building of more flexible GPD models capable of description of the existing and oncoming
DVCS and HMP data.

\section*{Acknowledgments}
The work of K.S. has been supported by
by the Russian Science Foundation (Grant No. 14-22-00281).

\end{document}